\documentclass[twocolumn,letterpaper,aps,prc,superscriptaddress,showpacs,nofootinbib,floatfix]{revtex4}

\usepackage{graphicx} 
\usepackage{hyperref} 
\usepackage{mathbbol} 

\begin{document}

\title{Suppression of away-side jet fragments with respect to the reaction plane
  in Au+Au collisions at $\sqrt{s_{_{NN}}}$ = 200 GeV}

\newcommand{\abilene}{Abilene Christian University, Abilene, Texas 79699, USA}
\newcommand{\acadsin}{Institute of Physics, Academia Sinica, Taipei 11529, Taiwan}
\newcommand{\banaras}{Department of Physics, Banaras Hindu University, Varanasi 221005, India}
\newcommand{\barc}{Bhabha Atomic Research Centre, Bombay 400 085, India}
\newcommand{\bnlcoll}{Collider-Accelerator Department, Brookhaven National Laboratory, Upton, New York 11973-5000, USA}
\newcommand{\bnlphys}{Physics Department, Brookhaven National Laboratory, Upton, New York 11973-5000, USA}
\newcommand{\caucr}{University of California - Riverside, Riverside, California 92521, USA}
\newcommand{\charlesczech}{Charles University, Ovocn\'{y} trh 5, Praha 1, 116 36, Prague, Czech Republic}
\newcommand{\chonbuk}{Chonbuk National University, Jeonju, 561-756, Korea}
\newcommand{\ciae}{China Institute of Atomic Energy (CIAE), Beijing, People's Republic of China}
\newcommand{\cns}{Center for Nuclear Study, Graduate School of Science, University of Tokyo, 7-3-1 Hongo, Bunkyo, Tokyo 113-0033, Japan}
\newcommand{\colorado}{University of Colorado, Boulder, Colorado 80309, USA}
\newcommand{\columbia}{Columbia University, New York, New York 10027 and Nevis Laboratories, Irvington, New York 10533, USA}
\newcommand{\czechtech}{Czech Technical University, Zikova 4, 166 36 Prague 6, Czech Republic}
\newcommand{\dapnia}{Dapnia, CEA Saclay, F-91191, Gif-sur-Yvette, France}
\newcommand{\debrecen}{Debrecen University, H-4010 Debrecen, Egyetem t{\'e}r 1, Hungary}
\newcommand{\elte}{ELTE, E{\"o}tv{\"o}s Lor{\'a}nd University, H - 1117 Budapest, P{\'a}zm{\'a}ny P. s. 1/A, Hungary}
\newcommand{\ewha}{Ewha Womans University, Seoul 120-750, Korea}
\newcommand{\fit}{Florida Institute of Technology, Melbourne, Florida 32901, USA}
\newcommand{\fsu}{Florida State University, Tallahassee, Florida 32306, USA}
\newcommand{\gsu}{Georgia State University, Atlanta, Georgia 30303, USA}
\newcommand{\hiroshima}{Hiroshima University, Kagamiyama, Higashi-Hiroshima 739-8526, Japan}
\newcommand{\ihepprot}{IHEP Protvino, State Research Center of Russian Federation, Institute for High Energy Physics, Protvino, 142281, Russia}
\newcommand{\illuiuc}{University of Illinois at Urbana-Champaign, Urbana, Illinois 61801, USA}
\newcommand{\instpasczech}{Institute of Physics, Academy of Sciences of the Czech Republic, Na Slovance 2, 182 21 Prague 8, Czech Republic}
\newcommand{\isu}{Iowa State University, Ames, Iowa 50011, USA}
\newcommand{\jinrdubna}{Joint Institute for Nuclear Research, 141980 Dubna, Moscow Region, Russia}
\newcommand{\jyvaskyla}{Helsinki Institute of Physics and University of Jyv{\"a}skyl{\"a}, P.O.Box 35, FI-40014 Jyv{\"a}skyl{\"a}, Finland}
\newcommand{\kek}{KEK, High Energy Accelerator Research Organization, Tsukuba, Ibaraki 305-0801, Japan}
\newcommand{\kfki}{KFKI Research Institute for Particle and Nuclear Physics of the Hungarian Academy of Sciences (MTA KFKI RMKI), H-1525 Budapest 114, POBox 49, Budapest, Hungary}
\newcommand{\korea}{Korea University, Seoul, 136-701, Korea}
\newcommand{\kurchatov}{Russian Research Center ``Kurchatov Institute", Moscow, Russia}
\newcommand{\kyoto}{Kyoto University, Kyoto 606-8502, Japan}
\newcommand{\labllr}{Laboratoire Leprince-Ringuet, Ecole Polytechnique, CNRS-IN2P3, Route de Saclay, F-91128, Palaiseau, France}
\newcommand{\lawllnl}{Lawrence Livermore National Laboratory, Livermore, California 94550, USA}
\newcommand{\losalamos}{Los Alamos National Laboratory, Los Alamos, New Mexico 87545, USA}
\newcommand{\lpc}{LPC, Universit{\'e} Blaise Pascal, CNRS-IN2P3, Clermont-Fd, 63177 Aubiere Cedex, France}
\newcommand{\lund}{Department of Physics, Lund University, Box 118, SE-221 00 Lund, Sweden}
\newcommand{\maryland}{University of Maryland, College Park, Maryland 20742, USA}
\newcommand{\mass}{Department of Physics, University of Massachusetts, Amherst, Massachusetts 01003-9337, USA }
\newcommand{\muenster}{Institut fur Kernphysik, University of Muenster, D-48149 Muenster, Germany}
\newcommand{\muhlenberg}{Muhlenberg College, Allentown, Pennsylvania 18104-5586, USA}
\newcommand{\myongji}{Myongji University, Yongin, Kyonggido 449-728, Korea}
\newcommand{\nagasaki}{Nagasaki Institute of Applied Science, Nagasaki-shi, Nagasaki 851-0193, Japan}
\newcommand{\newmex}{University of New Mexico, Albuquerque, New Mexico 87131, USA }
\newcommand{\nmsu}{New Mexico State University, Las Cruces, New Mexico 88003, USA}
\newcommand{\ornl}{Oak Ridge National Laboratory, Oak Ridge, Tennessee 37831, USA}
\newcommand{\orsay}{IPN-Orsay, Universite Paris Sud, CNRS-IN2P3, BP1, F-91406, Orsay, France}
\newcommand{\peking}{Peking University, Beijing, People's Republic of China}
\newcommand{\pnpi}{PNPI, Petersburg Nuclear Physics Institute, Gatchina, Leningrad region, 188300, Russia}
\newcommand{\riken}{RIKEN Nishina Center for Accelerator-Based Science, Wako, Saitama 351-0198, Japan}
\newcommand{\rikjrbrc}{RIKEN BNL Research Center, Brookhaven National Laboratory, Upton, New York 11973-5000, USA}
\newcommand{\rikkyo}{Physics Department, Rikkyo University, 3-34-1 Nishi-Ikebukuro, Toshima, Tokyo 171-8501, Japan}
\newcommand{\saispbstu}{Saint Petersburg State Polytechnic University, St. Petersburg, Russia}
\newcommand{\saopaulo}{Universidade de S{\~a}o Paulo, Instituto de F\'{\i}sica, Caixa Postal 66318, S{\~a}o Paulo CEP05315-970, Brazil}
\newcommand{\seoulnat}{Seoul National University, Seoul, Korea}
\newcommand{\stonybrkc}{Chemistry Department, Stony Brook University, SUNY, Stony Brook, New York 11794-3400, USA}
\newcommand{\stonycrkp}{Department of Physics and Astronomy, Stony Brook University, SUNY, Stony Brook, New York 11794-3400, USA}
\newcommand{\subatech}{SUBATECH (Ecole des Mines de Nantes, CNRS-IN2P3, Universit{\'e} de Nantes) BP 20722 - 44307, Nantes, France}
\newcommand{\tenn}{University of Tennessee, Knoxville, Tennessee 37996, USA}
\newcommand{\titech}{Department of Physics, Tokyo Institute of Technology, Oh-okayama, Meguro, Tokyo 152-8551, Japan}
\newcommand{\tsukuba}{Institute of Physics, University of Tsukuba, Tsukuba, Ibaraki 305, Japan}
\newcommand{\vandy}{Vanderbilt University, Nashville, Tennessee 37235, USA}
\newcommand{\waseda}{Waseda University, Advanced Research Institute for Science and Engineering, 17 Kikui-cho, Shinjuku-ku, Tokyo 162-0044, Japan}
\newcommand{\weizmann}{Weizmann Institute, Rehovot 76100, Israel}
\newcommand{\yonsei}{Yonsei University, IPAP, Seoul 120-749, Korea}
\affiliation{\abilene}
\affiliation{\acadsin}
\affiliation{\banaras}
\affiliation{\barc}
\affiliation{\bnlcoll}
\affiliation{\bnlphys}
\affiliation{\caucr}
\affiliation{\charlesczech}
\affiliation{\chonbuk}
\affiliation{\ciae}
\affiliation{\cns}
\affiliation{\colorado}
\affiliation{\columbia}
\affiliation{\czechtech}
\affiliation{\dapnia}
\affiliation{\debrecen}
\affiliation{\elte}
\affiliation{\ewha}
\affiliation{\fit}
\affiliation{\fsu}
\affiliation{\gsu}
\affiliation{\hiroshima}
\affiliation{\ihepprot}
\affiliation{\illuiuc}
\affiliation{\instpasczech}
\affiliation{\isu}
\affiliation{\jinrdubna}
\affiliation{\jyvaskyla}
\affiliation{\kek}
\affiliation{\kfki}
\affiliation{\korea}
\affiliation{\kurchatov}
\affiliation{\kyoto}
\affiliation{\labllr}
\affiliation{\lawllnl}
\affiliation{\losalamos}
\affiliation{\lpc}
\affiliation{\lund}
\affiliation{\maryland}
\affiliation{\mass}
\affiliation{\muenster}
\affiliation{\muhlenberg}
\affiliation{\myongji}
\affiliation{\nagasaki}
\affiliation{\newmex}
\affiliation{\nmsu}
\affiliation{\ornl}
\affiliation{\orsay}
\affiliation{\peking}
\affiliation{\pnpi}
\affiliation{\riken}
\affiliation{\rikjrbrc}
\affiliation{\rikkyo}
\affiliation{\saispbstu}
\affiliation{\saopaulo}
\affiliation{\seoulnat}
\affiliation{\stonybrkc}
\affiliation{\stonycrkp}
\affiliation{\subatech}
\affiliation{\tenn}
\affiliation{\titech}
\affiliation{\tsukuba}
\affiliation{\vandy}
\affiliation{\waseda}
\affiliation{\weizmann}
\affiliation{\yonsei}
\author{A.~Adare} \affiliation{\colorado}
\author{S.~Afanasiev} \affiliation{\jinrdubna}
\author{C.~Aidala} \affiliation{\mass}
\author{N.N.~Ajitanand} \affiliation{\stonybrkc}
\author{Y.~Akiba} \affiliation{\riken} \affiliation{\rikjrbrc}
\author{H.~Al-Bataineh} \affiliation{\nmsu}
\author{J.~Alexander} \affiliation{\stonybrkc}
\author{K.~Aoki} \affiliation{\kyoto} \affiliation{\riken}
\author{L.~Aphecetche} \affiliation{\subatech}
\author{Y.~Aramaki} \affiliation{\cns}
% ...
\collaboration{PHENIX Collaboration} \noaffiliation

\date{\today}
 
\begin{abstract}
  Pair correlations between large transverse momentum neutral pion
  triggers ($p_{T}$ = 4--7 GeV/$c$) and charged hadron partners
  ($p_{T}$ = 3--7 GeV/$c$) in central (0--20\%) and midcentral (20--60\%) 
  Au+Au
  collisions are presented as a function of trigger orientation with
  respect to the reaction plane.  The particles are at larger momentum
  than where jet shape modifications have been observed, and the
  correlations are sensitive to the energy loss of partons traveling
  through hot dense matter.  An out-of-plane trigger particle produces
  only $26\pm20\%$ of the away-side pairs that are observed opposite of
  an in-plane trigger particle.  In contrast, near-side jet fragments
  are consistent with no suppression or dependence on trigger
  orientation with respect to the reaction plane.  These observations
  are qualitatively consistent with a picture of little near-side parton
  energy loss either due to surface bias or fluctuations and increased
  away-side parton energy loss due to a long path through the
  medium.  The away-side suppression as a function of reaction-plane
  angle is shown to be sensitive to both the energy loss mechanism in
  and the space-time evolution of heavy-ion collisions.
\end{abstract}


\maketitle

\section{Introduction} 

Collisions of heavy nuclei at the Relativistic Heavy Ion Collider
have created matter with energy densities exceeding the
predicted threshold for deconfinement of color charge into a hot
dense plasma~\cite{ppg048}.  In this quark gluon plasma(QGP), quarks 
and gluons are not bound within hadronic states and the matter behaves 
collectively.  Comparisons with hydrodynamic simulations indicate
rapid thermalization of the colliding system into a hot dense nuclear 
medium.  The produced medium affords an opportunity to study the 
properties of a new phase
of quantum chromodynamics (QCD) in an extreme environment.

Hard scattering with large momentum exchange between 
partons in the incoming nuclei is well-described by
perturbative QCD (pQCD).  The scattered partons emerge 
back-to-back in azimuth in the plane transverse to the beam direction,
and fragment into a pair of correlated cones of high momentum particles,
 referred to as jets.  The study of jets and their hadronic fragments 
in heavy-ion collisions provides insight into the
properties of hot dense nuclear matter.
Measurements of single high transverse momentum ($p_T$)
particles~\cite{ppg054} and correlations between high-$p_T$
particles~\cite{starhighpt, ppg083, ppg106} have demonstrated that
the fast partons embedded in the produced medium dissipate a large amount
of their initial kinetic energy.

In this paper, we present angular correlations of
hadron pairs with both hadrons in the midrapidity range
$|\eta|<0.35$.  Fragments from the same jet form a peak 
at small relative azimuthal angle ($\Delta\phi$), i.e.  the near-side
peak.  Pairs composed of one fragment from each jet
will appear in an away-side peak at $\Delta\phi\sim\pi$.  Past
measurements~\cite{starhighpt, ppg083, ppg106} for hadrons $\gtrsim$ 5
GeV/$c$ have shown that the away-side correlations peak is suppressed 
relative to baseline measurements in $p$+$p$ collisions.  The
suppression of the away-side jet is a signature of parton energy loss
inside the medium.  The same measurements show that near-side jet
fragments at large momentum are not suppressed.
This feature of the data is understood to result
from the requirement of a large momentum particle in the final
state, which creates a bias towards small energy loss either by the
preferential selection of hard scatterings near the medium
surface~\cite{surfacebias} or due to fluctuations in energy
loss~\cite{elossfluct}.  

The detailed mechanism by which partons lose energy when passing
through a deconfined medium are not yet fully understood.  In pQCD
descriptions of the parton-medium interaction the predicted parton
energy loss should scale as the path length
squared~\cite{Dominguez:2008vd}.  In competing anti-de-Sitter
space/conformal field theory descriptions characterizing a
strongly coupled medium, the energy loss scales as the path length
cubed~\cite{Dominguez:2008vd}.  The variation in azimuthal angle of the
away-side jet suppression with respect to the reaction plane
($\phi_s$) is sensitive to the total amount of energy lost by the 
away-side parton along long paths (out-of-plane) or short paths
(in-plane) through the medium.  The degree to which the away-side jet
suppression varies will be determined in part by the path-length scale
of energy loss.

Single particle observables at high $p_{T}$, such as the nuclear
suppression with respect to the reaction plane ($R_{\rm AA}(\phi_s)$) or
the azimuthal anisotropy (i.e.  $v_{2}$)~\cite{ppg110}, are also
sensitive to this path-length variation of energy loss.  Current pQCD
calculations predict a lower $v_{2}$ than is found in the data and may
imply a larger than path length squared dependence to energy loss~\cite{pQCD1,pQCD2}.  The
reaction-plane dependence of the back-to-back jets provides an
additional test on the path length dependence in that the two particle 
observable selects a different distribution of hard-scattering
locations and should probe longer paths through the medium than single particle
observables.  The path-length dependence of both single and two
particle observables have already been studied through selection of the
collision centrality~\cite{ppg110,ppg083}.  However, centrality
selection varies not only the path length, but also other important
properties of the medium (e.g.  the overall energy density).  Selection
with respect to the reaction plane more directly varies the path
lengths, while leaving the other medium properties unchanged.

In addition to the uncertainties associated with the energy loss
mechanisms, many of the details within hydrodynamic simulations of
heavy-ion collisions have not been fully constrained and tested 
by
experiment.  For instance, one such uncertainty is the geometrical
description of the energy deposited by the colliding nuclei which
could contribute to the degree of away-side suppression variation with
respect to the reaction plane.  Two competing descriptions, the Glauber
model~\cite{glauber} and the Color Glass Condensate~\cite{cgc},
predict different azimuthal distributions of matter with respect to the reaction
plane.  Thus the two descriptions give
different in-plane and out-of-plane path lengths through the
medium.  These descriptions are also used as different starting points
to the hydrodynamic evolution of the medium.  Other model uncertainties
include, but are not limited to, the extent of geometry fluctuations,
the time required for thermalization into a hydrodynamic medium, the
characteristics of the phase transition to confined hadrons, and the conditions under
which those hadrons become free-streaming particles into the vacuum.
These ambiguities in the proper modeling of heavy-ion 
collisions can
result in significant uncertainty in the extracted properties of 
the medium, such as the shear viscosity~\cite{Luzum:2008cw}.

In midcentral collisions (the middle 20--60\% of the total cross section) 
the variation of the away-side suppression is
expected to be largest as the collision zone
is the most anisotropic.  In contrast, central collisions are much more
isotropic and so provide a sample of events with small
anisotropy expected in the away-side suppression.  For instance, in the
Glauber model, midcentral events will have a
root-mean-square thickness through the medium of 3.2 fm in the
in-plane direction versus 4.8 fm in the out-of-plane direction, 
which is a 50\%
variation in path length.  However, for central 0--20\% collisions, 
the path length through the medium varies from 5.0 fm in the in-plane direction
to 5.8 fm in the out-of-plane direction, which is a much smaller 16\% 
variation.  It is notable that the thickness
through the medium in midcentral collisions changes more with respect
to the reaction plane than it does between central and midcentral collisions
where the away-side suppression at large momentum is already known to
vary~\cite{ppg083}.  Also worth noting is that the largest 
thickness in midcentral collisions is comparable to the shortest
thickness in central collisions.

Any prediction for the away-side suppression with respect to the 
reaction plane will be a convolution of the energy loss and a 
description of the space-time evolution of the medium. In the limit 
where the medium is never fully opaque to fast partons and the energy 
lost by the typical parton is some fraction of its initial energy, the 
away-side suppression will increase with angle with respect to the 
reaction plane.  This results because the average path length through 
the medium of the recoil parton is longer when out of the reaction 
plane.  It is possible, in the extreme limit of a medium with a large 
opaque core and thin transparent corona, the away-side suppression 
could instead weaken as the trigger particle orientation varies from 
the in-plane to out-of-plane directions.  The weakening in the thin 
corona scenario results from two effects; a larger relative number of 
scattering centers producing a pair of back-to-back final-state 
particles in the out-of-plane direction, but also the variation of the 
trigger particle multiplicity by angle with respect to the reaction 
plane.  However, it is worth noting that a large core and thin corona 
is an extreme configuration.  Variations within more realistic models 
of the away-side suppression will be intermediate between these extreme 
scenarios.

\begin{figure}[t]
  \centering
  \includegraphics[width=0.8\linewidth]{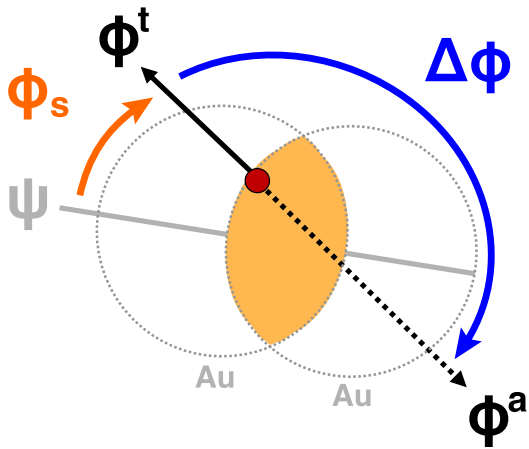}
  \caption{\label{fig:phiSdef} 
    (Color online) Definition of azimuthal angles.  Trigger and
    associated partner particles are measured at $\phi^t$ and
    $\phi^a$, respectively.  The trigger particle orientations are
    taken with respect to the reaction plane, $\phi_s = \phi^t -
    \psi$.  The relative azimuthal separation of the trigger particle
    and the associated partner particle is $\Delta\phi = \phi^t -
    \phi^a$.  
  }
\end{figure}

In this paper, we present azimuthal correlation measurements between
pairs of neutral pion trigger particles ($t$) within $p^t_T =$ 4--7
GeV/$c$ and charged hadron associated partner particles ($a$) within
$p^a_T =$ 3--4, 4--5, and 5--7 GeV/$c$.  These combinations of final-state
particle momentum ranges have previously been shown to be dominated by
jet fragmentation as they are above medium-induced two-particle
correlations which contribute significantly at lower
momenta~\cite{ppg083,ppg106}.  The low momentum structures (the
``ridge'' and ``shoulder'') 
may be the result of parton-medium interactions (e.g.~\cite{mach, Gubser:2007ga,largAng1, largAng2})
or global correlations from fluctuating initial
conditions~\cite{ps,Takahashi:2009na}.  These fluctuations have
substantially less impact at large pair momentum where the
background contribution becomes small.  In this study, as illustrated
in Fig.~\ref{fig:phiSdef}, using only
particles at large pair momentum, the 
away-side ($\Delta\phi\approx\pi$) suppression by trigger 
particle orientation with respect to the reaction plane ($\phi_s =
\phi^t - \psi$) is presented as a probe of both the mechanism of
parton energy loss and the space-time evolution of matter created by
the collisions of large nuclei.  

A previous measurement~\cite{Adams:2004wz} by the STAR collaboration 
for 20--60\% centrality between 4--6 GeV/$c$ trigger particles and 
$2<p^a_T<p^t_T$ GeV/$c$ partner particles for two $45^{\circ}$ wide 
in-plane and out-of-plane selections indicated an increased suppression 
of the away-side jet for the out-of-plane, but with little significance 
due to large underlying event subtraction uncertainties.  The new 
results presented in our paper have sufficient statistics to specify a 
trend in the away-side suppression in midcentral collisions at larger 
momentum where subtraction uncertainties are negligible.

\section{Experiment} 

The results presented here are based on 3.4 billion minimum-bias Au+Au
events recorded by the PHENIX detector in 2007.  Comparisons to $p$+$p$ 
collisions use previously published measurements from data recorded in 
2006~\cite{ppg106}.  Collision centrality was
determined by division into percentile of the integrated charge 
collected by beam-beam counters (BBC)~\cite{bbcref} located at
$|\eta|$ between 3.0 and 3.9.  The timing between the arrival of
charged particles in the north and south BBC was used to reconstruct
the event position along the collision axis ($z$-vertex), and to
restrict the event sample to $\pm30$ cm of the nominal interaction
point of the two beams.  The orientation of reaction-plane azimuthal
angle ($\psi$) is reconstructed event-by-event using the 
quadrupole component ($v_2$) of the charge in the Reaction Plane
(RXPN) detector~\cite{ppg098}, located at $|\eta|$ between 1.0 and
2.8.  The resolution of the RXPN detector is highest for midcentral
collisions
($\sim$20\%) where both the quadrupole component and the detector
occupancy are large.  The set of resolution corrections, $\Delta_{n}: n\in\{2,4,6,8\}$, for single
particle anisotropies, $v_{n}$, where:
\begin{eqnarray}
 v_{n} = \frac{v^{\rm obs}_n}{\Delta_n}
\label{eq:deltaDef1}
\end{eqnarray}
are estimated from correlations between the independent north ($\psi_{N}$) and
south ($\psi_{S}$) RXPN reaction-plane
reconstructions~\cite{Afanasiev:2009wq,ppg092}.  
A single fit parameter ($x$) is mapped into the
resolution corrections via: 
\begin{eqnarray}
  \Delta_{n} = \frac{1}{2} \sqrt{\pi} x e^{-\frac{x^2}{2}}
  \left( I_{\frac{n/2-1}{2}}\left(\frac{x^2}{2}\right) + I_{\frac{n/2-1}{2}}\left(\frac{x^2}{2}\right)\right)
\end{eqnarray}
The fit parameter is extracted from the correlations via: 
\begin{eqnarray}
  C(\psi_N-\psi_S) &=& \frac{1}{2}e^{-\frac{x^2}{2}}
  \left( \rule{0mm}{7.0mm} \right.
    \frac{2}{\pi} \left( \rule{0mm}{4.0mm}1+\frac{x^2}{2}\right) \nonumber \\
+ z \left( \rule{0mm}{4.0mm} I_0\left(z\right) \right.  &+& \left.  L_0\left(z\right)\rule{0mm}{4.0mm} \right) 
+ \frac{x^2}{2} \left( \rule{0mm}{4.0mm} I_1\left(z\right) + L_1\left(z\right) \right)
  \left.  \rule{0mm}{7.0mm} \right)
\label{eq:deltaDef2}
\end{eqnarray}
where
\begin{eqnarray}
  z = \frac{x^2}{2} \cos \left( \psi_N-\psi_S \right)
\end{eqnarray}

The set of functions, $I_{2k}$ and $L_{2k}$, are the even-ordered
modified Bessel functions and the modified Struve functions
respectively.  
The extracted values used to correct the measured second-order
azimuthal anisotropy, are $\Delta_2$ = 0.66(4) and 0.66(3) for 0--20\%
and 20--60\% collisions, respectively.  A 10\% systematic uncertainty in
0--5\% collisions and 5\% elsewhere accounts for non-flow contributions
to the resolution corrections~\cite{ppg098}.  The similar values are a
result of the peak in reaction-plane resolution appearing near 20\%
centrality.  A direct inspection of these reaction-plane distributions
for events containing a photon above 1 GeV/$c$ did not reveal
significant contributions from jets.  

Neutral pion trigger particles are reconstructed from photon clusters
measured by either lead-glass or lead-scintillator electromagnetic 
calorimeters (EMCal) in the two central arms of PHENIX, in total covering
$|\eta| < 0.35$ and $2\times90^{\circ}$ in
azimuth~\cite{ppg080}.  Clusters are subject to cuts based on the known
response of the EMCal, including noisy and low-response towers,
as well as shower shape cuts.  Neutral pions are identified
through the 2$\gamma$ decay channel by pairing all photons within an
event.  Incorrect pairings between photons create a broad combinatorial 
background under the $\pi^0$ mass peak.  This background is minimized
by requiring the reconstructed mass to lie near the $\pi^0$ mass
peak.  This requirement was 0.125--0.160 MeV/$c^2$ for central events, but
was relaxed to 0.120--0.165 MeV/$c^2$ in midcentral events where the
combinatorial background is lower.  Since combinatorial pairs are more often made
with the abundant photons found at low energy, the energy
asymmetry of the decay ($|E_1-E_2|/(E_1+E_2)$) was restricted to be less
than 0.5 for 0--5\% central events.  This was also relaxed slowly
for more peripheral events until all pairs with asymmetries less than
0.7 were accepted.  The tightness of the cuts was used to control the
rate of combinatorial pairings such that $\pi^0$ trigger particles
have a signal-to-background ratio averaged over the mass window of 4:1 in
central collisions and 10:1 in midcentral collisions.  

Charged hadron partner particles are reconstructed in the central arms
using the drift chambers (DC) with hit association requirements in two
layers of multi-wire proportional chambers with pad readout (PC1 and
PC3), achieving a momentum resolution, $\Delta p/p$, of $0.7\% \oplus 1.1\% p$
(GeV/$c$).  Only tracks with unambiguous and distinguishable DC and PC1 hit
information are used.  Projections of these tracks are required to
match a PC3 hit within a $\pm 2 \sigma$ proximity window to reduce
background from conversion and decay products.  A track association to
a signal in the Ring Imaging \v{C}erenkov detector is used to reject
electrons for partner selections below 5 GeV/$c$ where little signal
is produced by charged pions.

\section{Pair Analysis}

\begin{figure}[t]
  \centering
  \includegraphics[width=1.0\linewidth]{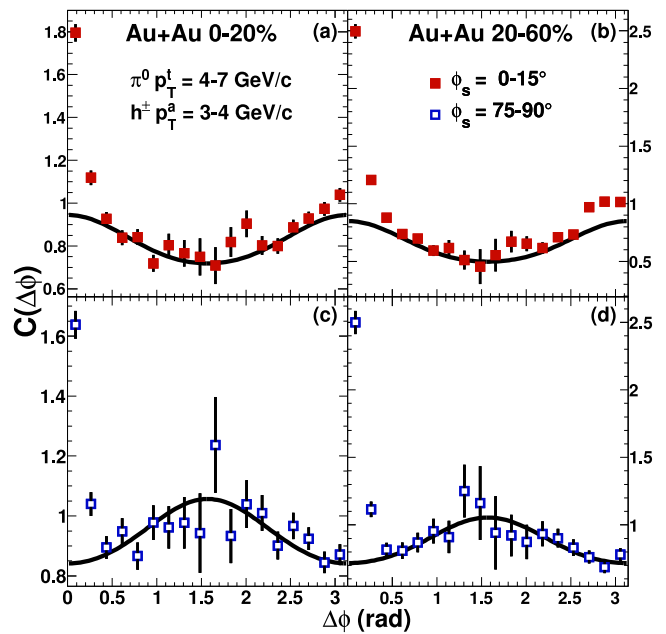}
  \caption{\label{fig:cfs}
    (Color online) Correlation functions for the most in-plane,
    $\phi_{s}$=0--15$^{\circ}$, (solid squares) and out-of-plane, 
$\phi_{s}$=75-90$^{\circ}$,
    (open squares) trigger $\pi^0$ orientations in central 0--20\% and
    midcentral 20--60\% collisions, left and right columns
    respectively, for 3--4 GeV/$c$ partner hadrons.  Expected underlying
    event contributions are shown as solid curves (see text for details).  
  } 
\end{figure}

Within an event, all pairs formed from $\pi^0$ trigger particles
($p^t_T$ = 4--7 GeV/$c$) and three sets of charged hadron associated partner particles
($p^a_T$ = 3--4, 4--5, 5--7 GeV/$c$) are measured.  Two centrality classes are used: a 
central selection of 0--20\% collisions and a midcentral selection of 20--60\%
collisions.  Trigger particles are separated into six $15^{\circ}$
bins in azimuthal angle with respect to the reaction plane, $\phi_s
= \phi^t-\psi$.  The angular resolution of the measured reaction plane,
at approximately $25^\circ$,
is larger than this binning; consequently, significant smearing
takes place between neighboring trigger orientation bins.  Pairs within PHENIX are
collected at different efficiencies due to the non-uniform 
central arm acceptance.  The relative pair efficiencies are
corrected by mixed pair distributions in which trigger and partner
particles are drawn from different events of the same class (bins of 5\% centrality,
5 cm $z$-vertex).  The resulting acceptance-corrected distributions are
reported as correlation functions, $C(\Delta\phi)$, which are defined as: 
\begin{equation}
  C(\Delta\phi) =  \frac{ \frac{d\mathbb{n}_{\rm same}^{ta}}{d\Delta\phi} }
                        { \frac{d\mathbb{n}_{\rm mix}^{ta}}{d\Delta\phi} }
                   \frac{ \int{ \frac{d\mathbb{n}_{\rm mix}^{ta}}{d\Delta\phi} d\Delta\phi} }
                  { \int{ \frac{d\mathbb{n}_{\rm same}^{ta}
                        }{d\Delta\phi} d\Delta\phi} }
\end{equation}
where $\mathbb{n}^{ta}$ is the number of measured pairs per event for
either the same or mixed events, as indicated.  Double-struck notation
($\mathbb{n}$) is used here to indicate measured quantities.  Representative 
correlation functions for in-plane and out-of-plane trigger particle 
orientations are shown in Fig.~\ref{fig:cfs}.  The full set of the
measured correlation functions used in this analysis is shown in
Figs.~\ref{fig:cfsall_00_20} and~\ref{fig:cfsall_20_60}.  Note that these
distributions are not corrected for reaction-plane resolution.  

These inclusive pairs are assumed to correlate in one of two ways.  (1) Two particles 
within the same event may correlate trivially by participation in the
same collision geometry.  These pairs produce an azimuthal angular correlation from the
the single particle anisotropy with respect to the reaction plane.
(2) Two particles may also correlate with each other via the same
hard-scattering process.  These particles will be fragments from the
same (di)jet.  To separate the jet particle pairs from the other
background pairs, the two-source assumption is expressed
as~\cite{ppg032}: 
\begin{eqnarray}
 C\left(\Delta\phi\right) &=& J\left(\Delta\phi\right)
\nonumber \\ 
&+& b_{0}\left(
1+\frac{\beta}{\alpha}\cos\left(2\Delta\phi\right)
+\frac{\gamma}{\alpha}\cos\left(4\Delta\phi\right)\right)
\label{eq:defCF}
\end{eqnarray}
where the jet contribution to the correlation function is contained in
$J(\Delta\phi)$.  The remaining harmonic terms 
describe the background contribution which is complicated by the
trigger particle binning with respect to the reaction plane.  The
background modulation coefficients ($\alpha,\beta,\gamma$) are
calculated via: 
\begin{eqnarray}
\alpha &=& 1 + 2 v^{t}_{2}\cos\left(2\phi_{s}\right)
\frac{\sin\left(2c\right)}{2c}\Delta_2 \nonumber \\ &+& 2
v^{t}_{2}\cos\left(4\phi_{s}\right)\Delta_{4} 
 \\
\beta &=& 2 v^{t}_{2} v^{a}_{2} + 2 v^{a}_{2} 
\left(1 + v^{t}_{4}\right)\cos\left(2\phi_{s}\right)
\frac{\sin\left(2c\right)}{2c}\Delta_2 
\nonumber \\
&+& 2 v^{t}_{2} v^{a}_{2} \cos\left(4\phi_{s}\right)
\frac{\sin\left(4c\right)}{4c}\Delta_4 
\nonumber \\
&+& 2 v^{a}_{2} v^{t}_{4} 
\cos\left(6\phi_{s}\right)\frac{\sin\left(6c\right)}{6c}\Delta_6
 \\
\gamma &=& 2 v^{t}_{4} v^{a}_{4} + 2 v^{a}_{4} \left(1 + v^{t}_{2}\right)
\cos\left(4\phi_{s}\right)\frac{\sin\left(4c\right)}{4c}\Delta_4
\nonumber \\
&+& 2 v^{t}_{2} v^{a}_{4} \left( \cos\left(2\phi_{s}\right)
\frac{\sin\left(2c\right)}{2c}\Delta_2 \right.
\nonumber \\
&+& \left.  \cos\left(6\phi_{s}\right)
\frac{\sin\left(6c\right)}{6c}\Delta_6 \right) 
\nonumber \\
&+& 2 v^{t}_{4} v^{a}_{4} \cos\left(8\phi_{s}\right)
\frac{\sin\left(8c\right)}{8c}\Delta_8
\label{eq:flowvariables}.
\end{eqnarray}
This description of the background accounts for the trigger particle
binning and reaction-plane resolution effects on the background
shape~\cite{flowsub}.  The trigger particle orientation appears
explicitly in terms of the bin center, $\phi_s$, and width, 
$c$.  Single particle anisotropy values, $v_2$ and $v_4$, were measured
by correlating the trigger and partner particles with respect to the
reaction plane, such that:  
\begin{eqnarray}
 C(\phi-\psi) = 1 &+& 2v^{\rm obs}_2\cos(2(\phi-\psi)) \nonumber \\ &+& 2v^{\rm obs}_4\cos(4(\phi-\psi))
\end{eqnarray}
where the observed anisotropies are corrected for the reaction-plane
resolution as described previously in Equation~\ref{eq:deltaDef1}.
Given sufficient detector resolution and narrowness of the trigger
particle orientation binning, the sign of the $\cos(2\Delta\phi)$ term
in Eq.~\ref{eq:defCF} will flip sign between in-plane and out-of-plane bins as 
shown in Fig.~\ref{fig:cfs}.  This effect is expected as selecting
out-of-plane trigger particles should decrease the likelihood of
finding a second background particle nearby.  The same is not true for
particles correlated via hard scattering.  Both the second- and
fourth-order anisotropy of the background correlations have 
been considered as the finite fourth-order contributions were
determined to be non-negligible for some trigger particle
orientations.  Likewise, higher-order terms in the reaction-plane
resolution correction are also included.   

The uncertainties on the reaction-plane resolution corrections
($\Delta_n$) and the observed anisotropies ($v^{\rm obs}_2$ ,$v^{\rm obs}_4$)
are propagated separately as they impact the away-side suppression
with respect to the reaction plane in characteristically different
ways.  The uncertainty in the reaction-plane resolution corrections is
fully correlated between trigger orientations.  For instance, this
uncertainty increases (or decreases) both the extracted in-plane and
out-of-plane jet yields at $\Delta\phi=\pi$.  However, the uncertainty
in the observed anisotropies is fully anti-correlated between trigger
orientations.  Thus, this uncertainty increases the extracted in-plane
yield while decreasing the out-of-plane yield (or vice versa).  At
large momentum, both of these subtraction uncertainties are small and
always dominated by other sources.  

The subtraction procedure was also examined for contamination of the jet
correlations by fakes in the charged tracking, which become
significant at large $p_T$.  The fake high $p_T$ tracks are present only in
the partner sample and are largely uncorrelated with trigger particle
for the partner $p_T$ presented here.  Thus the fake tracks, which are
already less influential in events with a high $p_T$ $\pi^0$, are
subtracted with other uncorrelated pairs as part of the background
contribution, so long as the anisotropies are measured with the same
particle cuts.  The subtraction robustness against tracking fakes at
high $p_{T}$ was checked by repeating the procedure with a 3$\sigma$
PC3 matching requirement.  

\begin{figure}[t]
  \centering
  \includegraphics[width=1.00\linewidth]{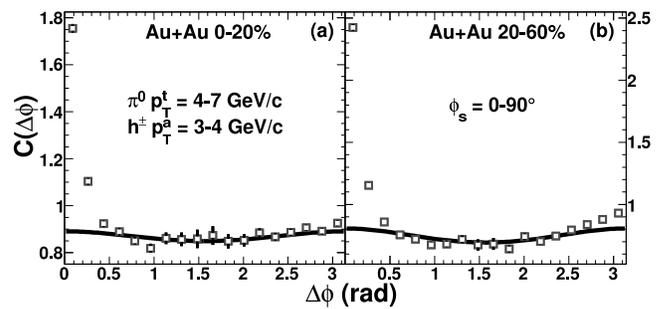}
  \caption{\label{fig:cfs_ave}
    Correlation functions for trigger $\pi^0$s averaged over
    all trigger orientations in central 0--20\% and midcentral 20--60\%
    collisions, left and right columns respectively, for 3--4 GeV/$c$
    partner hadrons.  Expected average underlying event contributions
    are shown as solid curves.
  } 
\end{figure}

By taking the trigger particle orientation as $\phi_s = \pi/4$, the
bin width as $c = \pi/2$, and by truncating higher than second-order
terms, the functional form of the background in Eq.~\ref{eq:defCF}
reduces to the $v^t_{2} \times v^a_{2}$ modulation 
used in previous trigger particle orientation averaged results such as
those found in~\cite{ppg083}.  This property is demonstrated in
Fig.~\ref{fig:cfs_ave} where the trigger particles from all
orientations are considered.  

The background level, $b_0$, is determined using the zero yield at
minimum (ZYAM) method~\cite{ppg032}.  At high-$p_T$, the well-separated near- and
away-side jets provide a large angular region at mid-$\Delta\phi$
angles with negligible jet signal.  This allows the ZYAM level to be
found with negligible bias and sufficient statistics despite the lower
efficiency PHENIX has for collecting pairs near $90^{\circ}$.  The ZYAM
uncertainty was estimated through simulation of the statistical
uncertainties as has been described in~\cite{abs}.

The jet contribution, $J(\Delta\phi)$, is then reported as a per-trigger
azimuthal yield such that:
\begin{eqnarray}
  \frac{1}{n^{t}}\frac{dn_{\rm jet}^{ta}}{d\Delta\phi} = \frac{1}{\epsilon^{a}} 
  \frac{\mathbb{n}_{\rm same}^{ta}}{\mathbb{n}^{t} \int{d\Delta\phi} }
  J(\Delta\phi).
\end{eqnarray}
The efficiency-corrected single particle and pair rates are $n^{t}$ and $n^{ta}$
respectively.  The single particle partner efficiency, $\epsilon^{a}$,
is estimated in simulations of detector acceptance and occupancy as
was done in~\cite{ppg106}.  By design, the trigger particle efficiency cancels in
the ratio.

\section{Results} 

\begin{figure*}[t]
\begin{minipage}{0.48\linewidth}
  \includegraphics[width=0.90\linewidth]{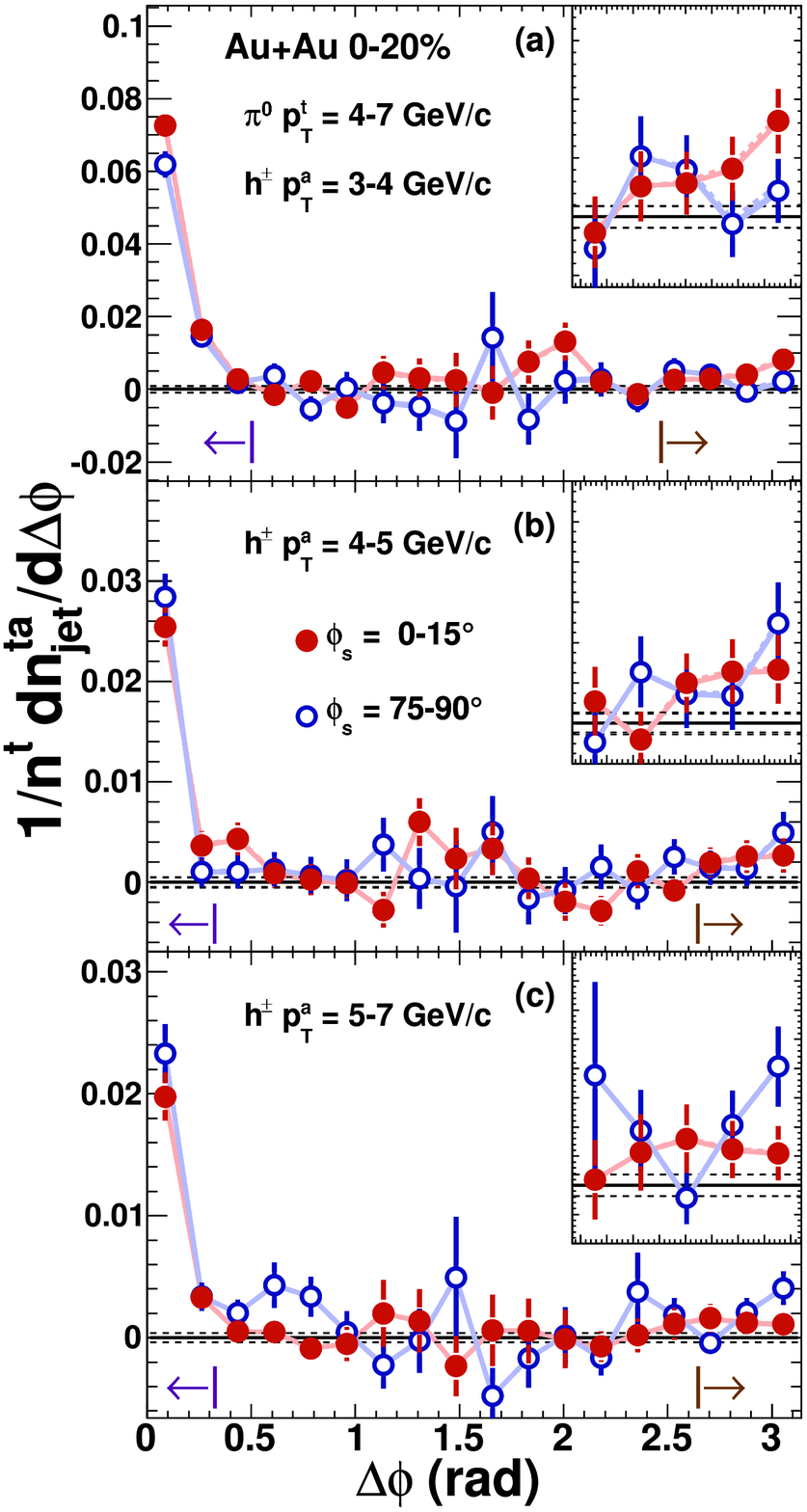}
  \caption{\label{fig:jfs_00_20} 
    (Color online) Per-trigger azimuthal jet yields for the most
    in-plane, $\phi_{s}$=0--15$^{\circ}$ (solid circles) and
    out-of-plane, $\phi_{s}$=75--90$^{\circ}$ (open circles) trigger
    particle selections in central 0--20\% collisions for various
    partner momenta.  Insets show away-side region on a zoomed
    scale.  Bars indicate statistical uncertainties.  Underlying event
    modulation systematic uncertainties are represented by bands
    through the points while the corresponding normalization
    uncertainties are shown as dashed lines around zero.  Near- and
    away-side jet yield integration windows are indicated with arrows.  
  }
\end{minipage}%
\hspace{0.5cm}
\begin{minipage}{0.48\linewidth}
  \includegraphics[width=0.90\linewidth]{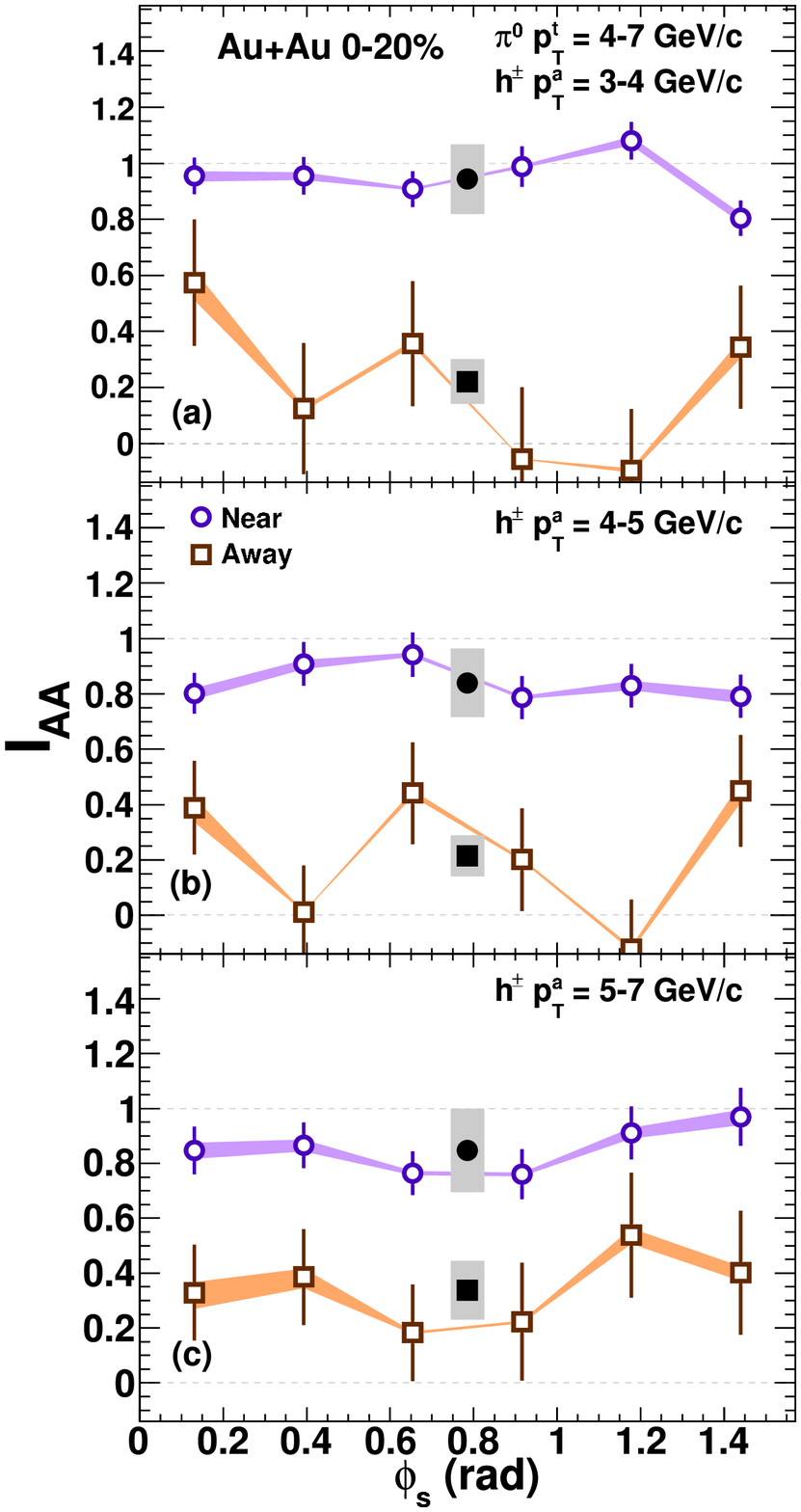}
  \caption{\label{fig:iaa_00_20}
    (Color online) Nuclear jet suppression factor, $I_{\rm AA}$, by
    angle with respect to the reaction plane, $\phi_s$, for near- and
    away-side angular selections, circles and squares respectively, in
    central 0--20\% collisions for various partner momenta.  Bars
    indicate statistical uncertainties.  The shaded band shows the
    systematic uncertainty on the reaction-plane resolution unsmearing
    correction.  Solid points show trigger particle angle averaged
    results and the global scale uncertainty.
  } 
\end{minipage}
\end{figure*}

Central events, 0--20\% collisions, are analyzed as a cross check
against experimental artifacts in midcentral collisions since they
have a smaller away-side jet yield.  Thus, the central events should 
exhibit a smaller trigger particle angle variation, 
require a larger reaction-plane resolution correction, 
a larger event correlation subtraction, and have increased
background in $\pi^0$ identification.  Representative per-trigger
azimuthal yields in central collisions for each of the partner
momentum selections for the most in-plane and most out-of-plane trigger
particle selections are shown in Fig.~\ref{fig:jfs_00_20}.
Figure~\ref{fig:jfsall_00_20} shows the full set of the measured 
per-trigger azimuthal yields used in this analysis for central 
collisions.  The most in-plane and most out-of-plane trigger-particle 
orientations select the shortest and longest average path lengths 
through the medium, respectively, and thus may be expected to 
have the maximimum differences.

On the near-side, a jet
distribution is clearly observed for each selection.  A direct
comparison between the most in-plane and most out-of-plane trigger 
shows no significant variation.  The
measurement at mid-$\Delta\phi$ demonstrates the good agreement
resulting from correct description of the underlying event
correlations.  On the away-side, the jet yield is small due to 
medium suppression and the statistical precision suffers once the pairs
are divided among the various trigger particle orientations.  
No evidence of experimental artifacts such as over-subtraction or 
incorrect description of the background is seen, despite the
challenging analysis environment present in central collisions.

\begin{figure*}[ht]
\begin{minipage}{0.48\linewidth}
  \includegraphics[width=0.90\linewidth]{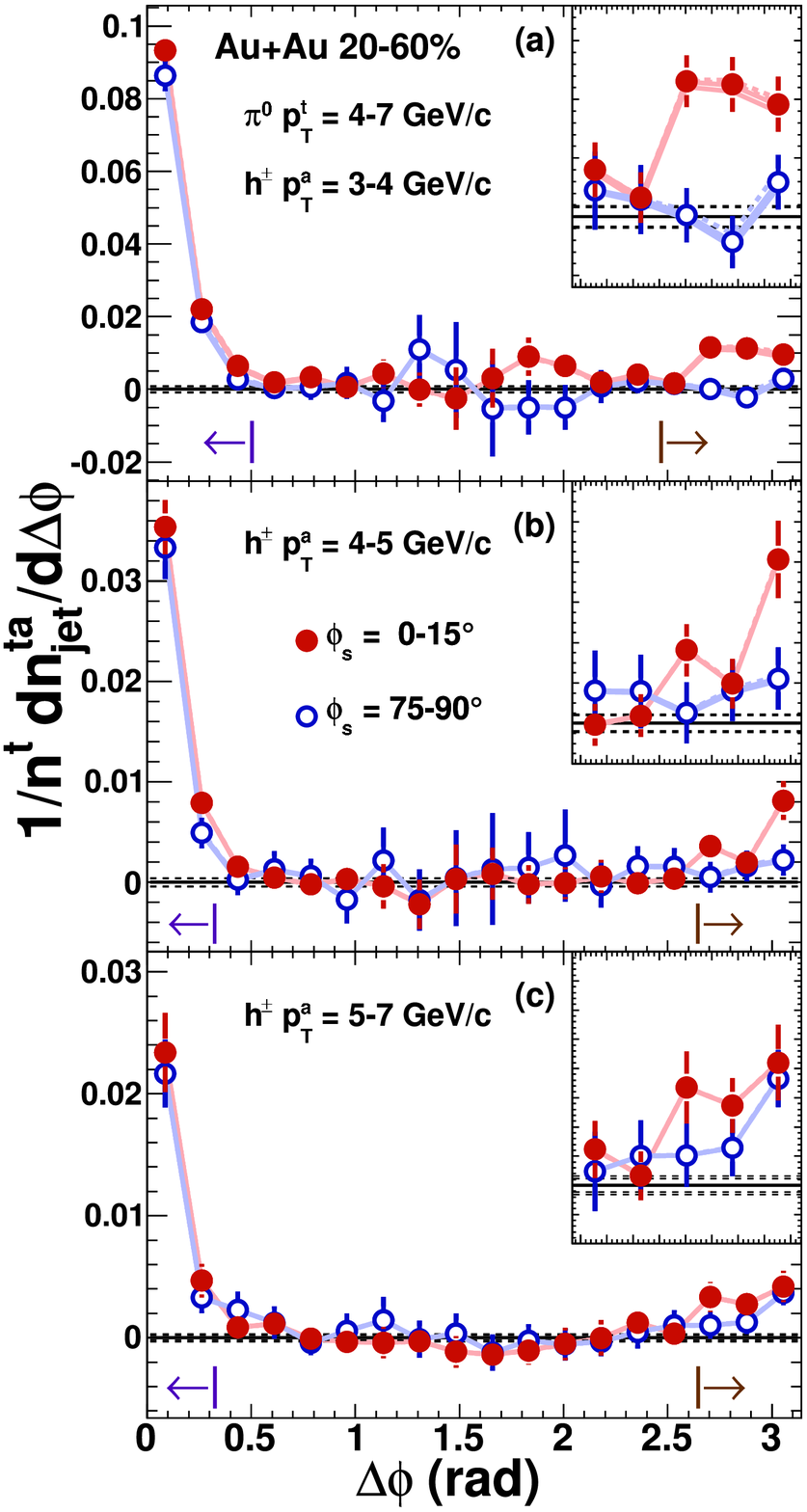}
  \caption{\label{fig:jfs_20_60}
    (Color online) Per-trigger azimuthal jet yields for the most
    in-plane, $\phi_{s}$=0--15$^{\circ}$, (solid circles) and
    out-of-plane, $\phi_{s}$=75--90$^{\circ}$, (open circles) trigger
    particle selections in midcentral 20--60\% collisions for various
    partner momenta.  Insets show away-side region on a zoomed
    scale.  Bars indicate statistical uncertainties.  Underlying event
    modulation systematic uncertainties are represented by bands
    through the points while the corresponding normalization
    uncertainties are shown as dashed lines around zero.  Near- and
    away-side jet yield integration windows are indicated with arrows.  
} 
\end{minipage}%
\hspace{0.5cm}
\begin{minipage}{0.48\linewidth}
  \includegraphics[width=0.90\linewidth]{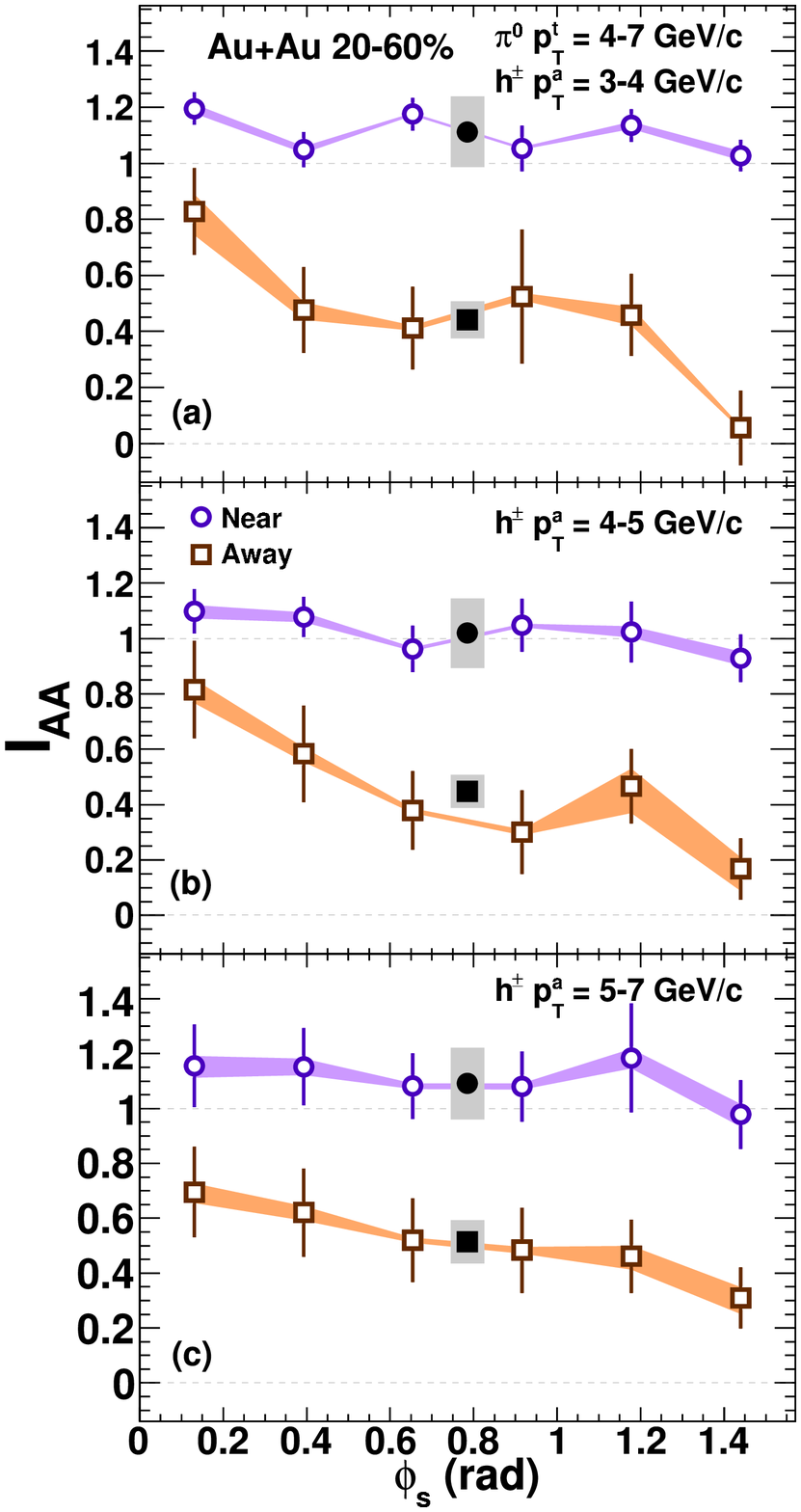}
  \caption{\label{fig:iaa_20_60}
    (Color online) Nuclear jet suppression factor, $I_{\rm AA}$, by
    angle with respect to the reaction plane, $\phi_s$, for near- and
    away-side angular selections, circles and squares respectively, in
    midcentral 20--60\% collisions for various partner momenta.  Bars
    indicate statistical uncertainties.  The shaded band shows the
    systematic uncertainty on the reaction-plane resolution unsmearing
    correction.  Solid points show trigger particle angle averaged
    results and the global scale uncertainty.
  } 
\end{minipage}
\end{figure*}

Integrated near- and away-side per-trigger yields ($Y$) are calculated
within angular $\Delta\phi$ windows, as indicated in
Fig.~\ref{fig:jfs_00_20}, approximating the $2\sigma$ width of the
jet distributions measured in the trigger particle orientation
averaged results.  The near-side azimuthal integration windows are
$\Delta\phi < \pi/9$ ($< 3\pi/18$) for $p^a_T > 4$ GeV/$c$ ($< 4$
GeV/$c$).  Similarly, the away-side azimuthal integrations windows are
$\pi-\Delta\phi < 3\pi/18$ ($< 2\pi/9$) for $p^a_T > 4$ GeV/$c$ ($<
4$ GeV/$c$).  Use of these windows corresponds to an assumption 
that the jet distributions do not widen significantly at high $p_T$, 
as a function of the trigger particle orientation with respect to the 
reaction plane.  This
assumption is supported by the absence of significant centrality
dependence in jet correlation widths ($\lesssim20\%$) for particles at high 
$p_T$~\cite{ppg106}.  Within statistical uncertainties a constant jet
width is consistent with the data.  Integrated yields as a function
of trigger particle orientation for both the near- and away-side are 
then corrected for the reaction-plane resolution.  The resolution 
correction is applied such that: 
\begin{eqnarray}
  Y(\phi_s) =
  \frac{1+2\left(v^{{\rm obs},Y}_{2}/\Delta_2\right)\cos\left(2\phi_{s}\right)}
  {1+2 v^{{\rm obs},Y}_{2}\cos\left(2\phi_{s}\right)} Y_{\rm meas}(\phi_s).  
\end{eqnarray}
where $Y$ and $Y_{\rm meas}$ are the corrected and uncorrected
yields, respectively.  The value of $v^{{\rm obs},Y}_{2}$ is the observed
second-order anisotropy of integrated per-trigger yield with respect to the reaction plane
and is determined by fitting the trigger particle orientation dependence of
each $Y_{\rm meas}(\phi_s)$ measurement individually.  This procedure is
the similar to the correction of reaction-plane resolution on single
particles, here applied to integrated per-trigger pair yields.

The corrected per-trigger yields ($Y$) are reported as the nuclear
jet suppression with respect to $p$+$p$ collisions, $I_{\rm AA}
= Y_{\rm{A}+\rm{A}}/Y_{p+p}$.  The result for central collisions is shown in
Fig.~\ref{fig:iaa_00_20}.  The variation of the fit used in the
resolution correction is the dominant source of $\phi_s$-correlated
uncertainty, having larger impact than the insignificant event
anisotropy uncertainties.  In the case of zero signal variation with
reaction plane orientation, the correction becomes completely
correlated with statistical scatter in the uncorrected
measurement.  Thus, the $\phi_s$-correlated systematic uncertainty from
the resolution correction is conservatively treated as correlated with
the statistical uncertainty when computing the final significance of
the measured trends.  For the same reason, this source of systematic
uncertainty has little correlation between the centrality and momentum
selections.   

For central events the near-side suppression is consistent with a constant as a function of
$\phi_s$ within the statistical and $\phi_s$-correlated systematic
uncertainties.  The values are also consistent with no suppression when
considering the global scale uncertainty that appears on the trigger
particle orientation averaged $I_{\rm AA}$.  On the away-side, there
is significant suppression in central events, as evidenced by the
trigger particle averaged $I_{\rm AA}$, but the statistical precision
with which to determine the $\phi_s$ variation is limited.

Mid-central events, 20--60\% collisions, have greater eccentricity and
could be expected to show correspondingly larger trigger particle
orientation dependence due to path-length variation through the
collision zone.  The same set of representative per-trigger azimuthal
yields is shown in Fig.~\ref{fig:jfs_20_60} for the midcentral
selection.  The full set at midcentrality is shown in
Fig.~\ref{fig:jfsall_20_60}.  Again, the near-side jets for the most
in-plane and most out-of-plane trigger particle orientations are
consistent with each other, a direct indication of little variation
with respect to the reaction plane.  The mid-$\Delta\phi$ are also in
agreement with zero, as before, further demonstrating that the underlying
event flow correlations are well described by
Equations~\ref{eq:defCF}-\ref{eq:flowvariables}.  In contrast to the
near-side, the away-side measurements (see insets in
Fig.~\ref{fig:jfs_20_60}) change between the in-plane and
out-of-plane trigger particle orientations with the latter having
consistently smaller yield for all partner momenta.

The integrated near- and away-side per-trigger jet yields for
midcentral collisions are shown in Fig.~\ref{fig:iaa_20_60}.  The
near-side jet is essentially flat, with negligible suppression
($I_{\rm AA}(\phi_s) = 1$).  The away-side jet yield is increasingly
suppressed with increasing $\phi_s$.  This falling trend results in
only small associated particle yield remaining for out-of-plane 
trigger particle orientations.

\begin{figure}[tb]
  \includegraphics[width=1.00\linewidth]{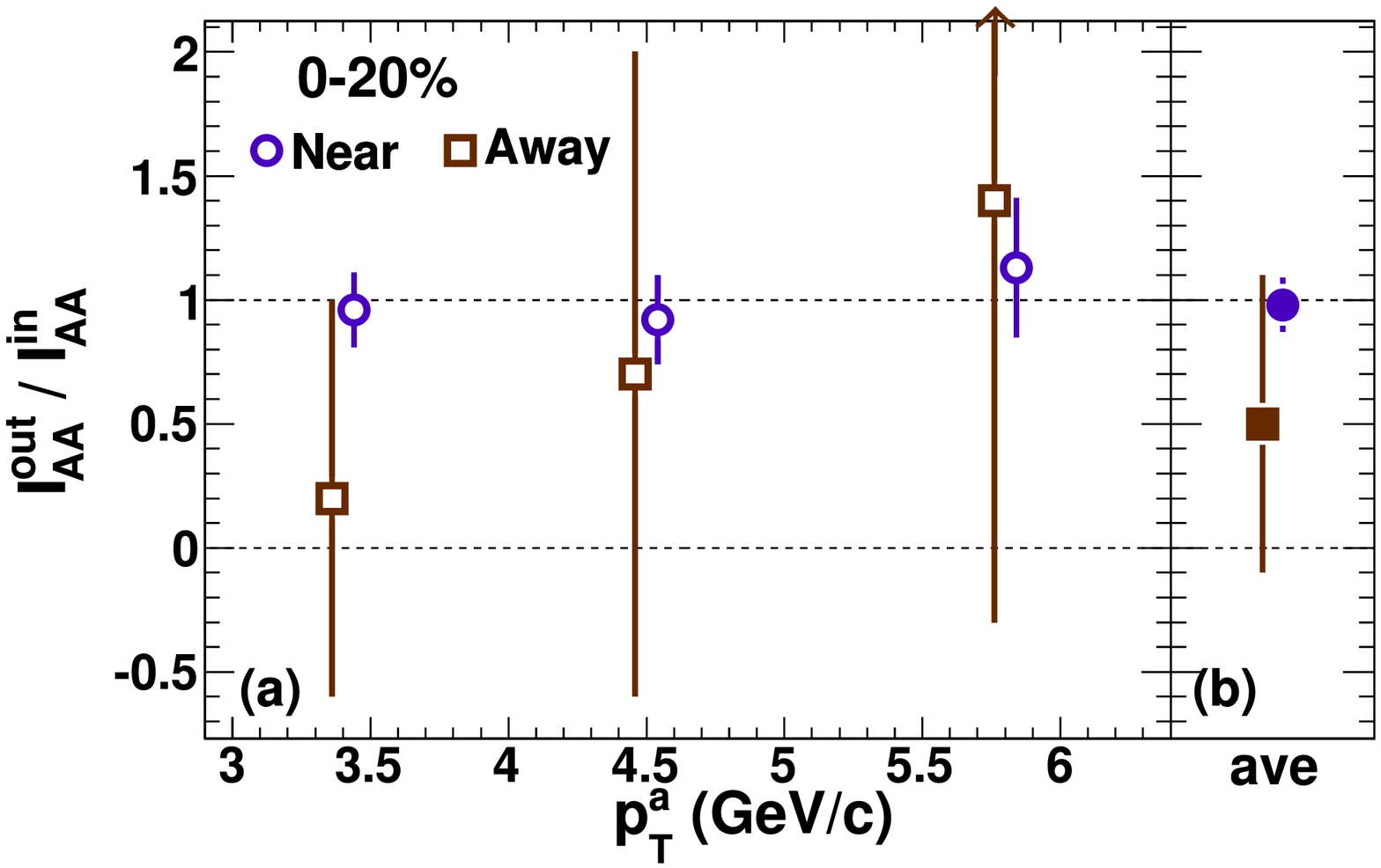}
  \caption{\label{fig:iaaratios_00_20}
  (Color online) Angle with respect to the reaction-plane dependence of the nuclear
  suppression factor, $I_{\rm AA}$, expressed as the ratio between
  in-plane and out-of-plane trigger particles from fits to the data in
  central 0--20\% collisions.  The bars represent total uncertainty
  taking into account the correlations between sources (see text for details).   
  } 
\end{figure}
\begin{figure}[tb]
  \includegraphics[width=1.00\linewidth]{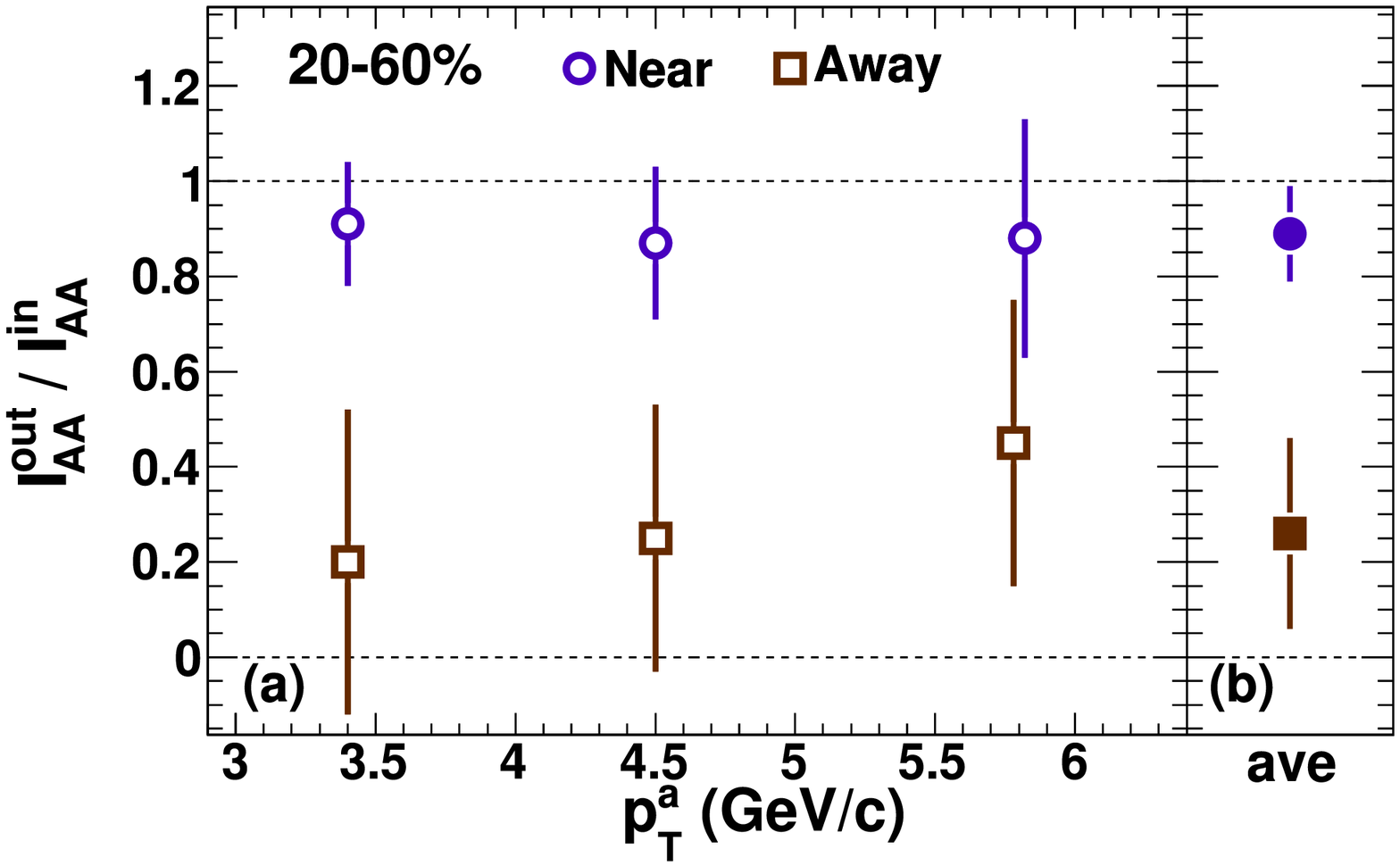}
  \caption{\label{fig:iaaratios_20_60}
  (Color online) Angle with respect to the reaction-plane dependence of the nuclear
  suppression factor, $I_{\rm AA}$, expressed as the ratio between
  in-plane and out-of-plane trigger particles from fits to the data in
  midcentral 20--60\% collisions.  The bars represent total uncertainty
  taking into account the correlations between sources (see text for details).
  } 
\end{figure}

\begin{table*}[t]
\caption{\label{tab:iaa}
  Angle with respect to the reaction-plane dependence of the nuclear
  suppression factor, $I_{\rm AA}$, expressed as the ratio between in-plane and
  out-of-plane trigger particles from linear and cosine fits to the data (see
  text for details).  The total uncertainty taking into account the
  correlations between sources is reported.  
}
\begin{ruledtabular} \begin{tabular}{cccccccccccc}
\multicolumn{2}{c}{Selection} & \multicolumn{5}{c}{Near-side} & \multicolumn{5}{c}{Away-side} \\
\multicolumn{2}{c}{} & \multicolumn{2}{c}{linear}  &
\multicolumn{2}{c}{cosine} & \multicolumn{1}{c}{average} & \multicolumn{2}{c}{linear}  &
\multicolumn{2}{c}{cosine} & \multicolumn{1}{c}{average} \\
Cent & $p^{a}_{T}$ & $I^{\rm out}_{\rm AA}/I^{\rm in}_{\rm AA}$ & $\chi^2/dof$ &
$I^{\rm out}_{\rm AA}/I^{\rm in}_{\rm AA}$ & $\chi^2/dof$ & & $I^{\rm out}_{\rm
  AA}/I^{\rm in}_{\rm AA}$ & $\chi^2/dof$ &
$I^{\rm out}_{\rm AA}/I^{\rm in}_{\rm AA}$ & $\chi^2/dof$ &  \\ 
\hline
0--20\% & 3--4 & $0.95 \pm 0.15$ & 9.5/4 & $0.96 \pm 0.15$ &  10.0/4 & $0.96 \pm 0.15$ & $0.1 \pm 0.7$ & 5.0/4 & $0.2 \pm 0.8$ & 5.1/4 & $0.2 \pm 0.8$\\
       & 4--5 & $0.92 \pm 0.18$ & 3.0/4 & $0.92 \pm 0.16$ &  3.0/4  & $0.92 \pm 0.18$ & $0.7 \pm 1.3$ & 9.0/4 & $0.6 \pm 1.2$ & 8.7/4 & $0.7 \pm 1.3$\\
       & 5--7 & $1.15 \pm 0.30$ & 3.1/4 & $1.10 \pm 0.26$ &  3.3/4  & $1.13 \pm 0.28$ & $1.5 \pm 2.0$ & 2.0/4 & $1.3 \pm 1.4$ & 1.8/4 & $1.4 \pm 1.7$\\
       & 3--7 &      ---        &  ---  &      ---        &   ---   & $0.98 \pm 0.11$ &      ---      &  ---  &      ---      &   --- & $0.5 \pm 0.6$\\
\hline
20--60\% & 3--4 & $0.90 \pm 0.14$ & 5.0/4 & $0.92 \pm 0.12$ & 5.5/4  & $0.91 \pm 0.13$ & $0.15 \pm 0.25$ & 4.0/4 & $0.25 \pm 0.38$ & 5.5/4 & $0.20 \pm 0.32$\\
        & 4--5 & $0.85 \pm 0.17$ & 1.2/4 & $0.88 \pm 0.15$ & 1.5/4  & $0.87 \pm 0.16$ & $0.20 \pm 0.20$ & 3.0/4 & $0.30 \pm 0.35$ & 4.0/4 & $0.25 \pm 0.28$\\
        & 5--7 & $0.88 \pm 0.28$ & 0.5/4 & $0.88 \pm 0.21$ & 0.7/4  & $0.88 \pm 0.25$ & $0.40 \pm 0.30$ & 0.3/4 & $0.50 \pm 0.30$ & 0.5/4 & $0.45 \pm 0.30$\\
        & 3--7 &     ---         &  ---  &      ---        &  ---   & $0.89 \pm 0.10$ &      ---        &  ---  &      ---        &   --- & $0.26 \pm 0.20$\\
\end{tabular} \end{ruledtabular}
\end{table*}

In order to quantify the variation and significance of the trigger
particle orientation dependencies shown in Figs.~\ref{fig:iaa_00_20}
and~\ref{fig:iaa_20_60}, the ratio of the out-of-plane to in-plane
suppression ($I^{\rm out}_{\rm AA}/I^{\rm in}_{\rm AA}$) is constructed.  In
the ratio, the global scale uncertainties on each measurement
cancel.  The $I_{\rm AA}$ values at $\phi_s$ = $0^{\circ}$
($I^{\rm in}_{\rm AA}$) and at $90^{\circ}$ ($I^{\rm out}_{\rm AA}$) are
estimated by both linear and flow-like cosine fits to the trigger
particle angle measurements and evaluation at these angles.  The
reported ratios are therefore independent of the chosen binning with
respect to the reaction plane and the values do not rely heavily on
the assumed functional form of the dependence.  The best-fit was
determined by $\chi^2$ minimization in which:
\begin{equation}
\tilde{\chi}^{2} = \sum \frac{\left(y_{i} + \epsilon_{sys}\sigma_{sys,i} - f\left(\phi_{s}\right)\right)^{2}}{\tilde{\sigma}^{2}_{i}\left(\epsilon_{sys}\right)} + \epsilon^{2}_{sys}
\end{equation}
where $\epsilon_{sys}$ is $\pm1$ for the $\pm1\sigma_{sys}$ variation of the
$\phi_s$-correlated systematic error~\cite{ppg079}.  As discussed above, the
systematic uncertainty is conservatively treated as fully correlated
with the statistical uncertainty.  The difference between linear and
cosine fits provides only a small source of additional uncertainty due to
the unknown true functional form.
The resulting values of $I^{\rm out}_{\rm AA}/I^{\rm in}_{\rm AA}$ and the total
uncertainty are shown in Figs.~\ref{fig:iaaratios_00_20}
and~\ref{fig:iaaratios_20_60}.  
The average value of
$I^{\rm out}_{\rm AA}/I^{\rm in}_{\rm AA}$ across partner
momentum is constructed by weighting the individual measurements by
the $p$+$p$ per-trigger yields~\cite{ppg106}.  In general, the data are
well fit by both the linear and cosine functions, giving reasonable
$\chi^2$.  No evidence is seen for systematic deviations from either fit within the sizable statistical
uncertainties and both forms give similar goodness of fit values.  
These values appear along with the $I^{\rm out}_{\rm AA}/I^{\rm in}_{\rm AA}$ ratios in Table~\ref{tab:iaa}.

For both central and midcentral collisions, the near-side jet yield
is independent of trigger particle orientation with respect to the
reaction plane within one standard deviation of the experimental
uncertainties.  These measurements are consistent with surface bias of
the hard scattering center created by the requirement of a trigger
particle and a resulting short path length through the collision zone
traversed by the near-side parton.  Central collisions have
insufficient statistics to determine the away-side variation.

However in midcentral collisions where the expectation of surface
bias would lead to a large variation in the path length traversed by
the away-side parton, the measurements show a significant falling
trend with increasing trigger particle angle with respect to the
reaction plane.  The suppression of away-side jet
fragments in the out-of-plane direction is larger than in the
in-plane direction, the out-of-plane away-side jet peak having only
$(26\pm20)\%$ of the yield of the in-plane direction.  Thus the large
variation by angle with respect to the reaction plane is
significant.  Assuming the modulation to be flow-like (dominated
by the second-order variation), the suppression pattern
implies $v_2^{I_{\rm AA}} = 0.29^{+0.15}_{-0.11}$.  As the midcentral
away-side measurements are consistent between $p^a_T$ selections
within the stated uncertainties, the hint of a rising trend in
$p^a_T$ is not significant.  The values
quoted here are consistent with those previously measured
in~\cite{Adams:2004wz} and provide a factor four better constraint in the
$I^{\rm out}_{\rm AA}/I^{\rm in}_{\rm AA}$ ratio.  

Recent single particle measurements of azimuthal anisotropy at high
$p_T$ ($6-9$ GeV/$c$) found that $v_{2} = 0.13 \pm 0.01 \pm
0.01$~\cite{ppg110}.  Thus, the away-side per-trigger yields at high
$p_T$ favor an anisotropy larger than that measured for the single
particles.  However, the difference is marginal and additional
measurements will be needed to confirm.  

\begin{figure}[tb]
  \includegraphics[width=1.00\linewidth]{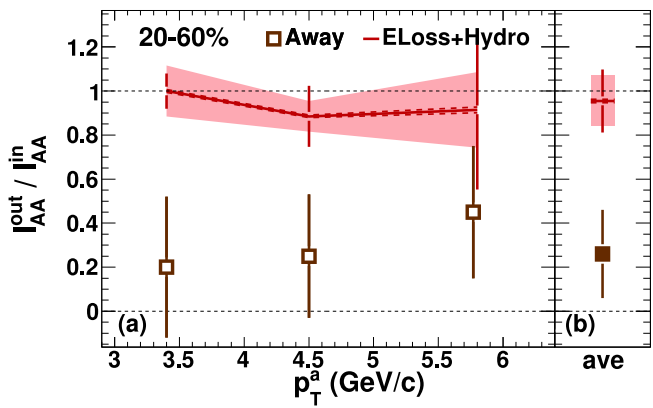}
  \caption{\label{fig:theory}
  (Color online) Away-side $I^{\rm out}_{\rm AA}/I^{\rm in}_{\rm AA}$ ratio for
  midcentral, 20--60\% collisions, from Fig.~\ref{fig:iaaratios_20_60}.  The solid line
  shows the results from an energy loss calculation~\cite{renkeloss,renkelossPC} using two
  hydrodynamic evolution models~\cite{dukehydro,jhydro}.  The shaded
  band shows the uncertainty that results from the selection of a
  particular hydrodynamic evolution; the lower extent
  covering~\cite{dukehydro} and the upper
  covering~\cite{jhydro}.  Dotted lines show the uncertainty from the
  initial event geometry (Glauber or CGC) as calculated
  within~\cite{dukehydro}.   
  } 
\end{figure}

Shown in Fig.~\ref{fig:theory} are the
results of a Monte-Carlo energy loss calculation from T.~Renk~\cite{renkeloss,renkelossPC} using the
time-space evolution provided by two different hydrodynamic
simulations~\cite{dukehydro,jhydro} and two initial state
descriptions, Glauber and CGC.  These particular combinations of a jet
energy loss model and collision evolution together predict less variation in
the away-side suppression with respect to the reaction plane than is
witnessed by the data.  Variation of the initial geometry description
within~\cite{dukehydro} between Glauber and CGC produces only small
changes in the extracted $I_{\rm AA}$ out-of-plane to in-plane ratio,
indicating limited sensitivity to this model parameter of the reaction
plane dependent dijet observable.
However, other model parameters that vary between the two
hydrodynamic models (such as the thermalization time and freeze-out
temperature) were found to impact the away-side suppression anisotropy
to a greater degree, indicating sensitivity to simulation parameters
that are not well-constrained by other measurements.  Consequently,
these data warrant more detailed study with various energy loss models, 
and also different space-time evolution models.  

\section{Summary} 

We have shown that away-side jet fragment suppression increases
substantially with increasing angle with respect to the reaction plane
in midcentral Au+Au collisions at $\sqrt{s_{_{NN}}} = 200$ GeV.  The away-side
yield in the out-of-plane orientation is reduced by a factor of $\sim4$
relative to the in-plane direction.  In contrast, the measured
near-side $I_{\rm AA}$ is reaction plane independent, and consistent 
with no suppression.  These
results directly show that the energy lost by fast partons in the hot
nuclear medium increases as their paths through the medium become
long.  A theoretical description of these experimental data
implementing an energy loss formalism and a time-space evolution of
the collision should be sought in union with other experimental
quantities; such as $R_{\rm AA}$, $I_{\rm AA}$, and
$R_{\rm AA}$($\phi_s$)~\cite{ppg054, ppg083, ppg090, ppg092,
  ppg106}.  Energy loss formalisms that have successfully described the
large momentum $R_{\rm AA}$ and $I_{\rm AA}$ may be paired with a particular
time-space evolution in also describing the $\phi_s$ dependence of
these same quantities.  As shown for the combination above, the data
presented here disagree with the present calculations.  These data
should play an important role in constraining 
simulations of the space-time evolution of heavy-ion collisions and
the subsequent extraction of medium properties.


\begin{acknowledgments}


We thank the staff of the Collider-Accelerator and Physics
Departments at Brookhaven National Laboratory and the staff of
the other PHENIX participating institutions for their vital
contributions.  We acknowledge support from the 
Office of Nuclear Physics in the
Office of Science of the Department of Energy,
the National Science Foundation, 
a sponsored research grant from Renaissance Technologies LLC, 
Abilene Christian University Research Council, 
Research Foundation of SUNY, 
and Dean of the College of Arts and Sciences, Vanderbilt University 
(U.S.A),
Ministry of Education, Culture, Sports, Science, and Technology
and the Japan Society for the Promotion of Science (Japan),
Conselho Nacional de Desenvolvimento Cient\'{\i}fico e
Tecnol{\'o}gico and Funda\c c{\~a}o de Amparo {\`a} Pesquisa do
Estado de S{\~a}o Paulo (Brazil),
Natural Science Foundation of China (People's Republic of China),
Ministry of Education, Youth and Sports (Czech Republic),
Centre National de la Recherche Scientifique, Commissariat
{\`a} l'{\'E}nergie Atomique, and Institut National de Physique
Nucl{\'e}aire et de Physique des Particules (France),
Ministry of Industry, Science and Tekhnologies,
Bundesministerium f\"ur Bildung und Forschung, Deutscher
Akademischer Austausch Dienst, and Alexander von Humboldt Stiftung (Germany),
Hungarian National Science Fund, OTKA (Hungary), 
Department of Atomic Energy and Department of Science and Technology (India),
Israel Science Foundation (Israel), 
National Research Foundation and WCU program of the 
Ministry Education Science and Technology (Korea),
Ministry of Education and Science, Russia Academy of Sciences,
Federal Agency of Atomic Energy (Russia),
VR and the Wallenberg Foundation (Sweden), 
the U.S.  Civilian Research and Development Foundation for the
Independent States of the Former Soviet Union, 
the US-Hungarian Fulbright Foundation for Educational Exchange,
and the US-Israel Binational Science Foundation.

\end{acknowledgments}

\appendix*
\section{Appendix}

\begin{figure*}[ht]
  \includegraphics[width=0.90\linewidth]{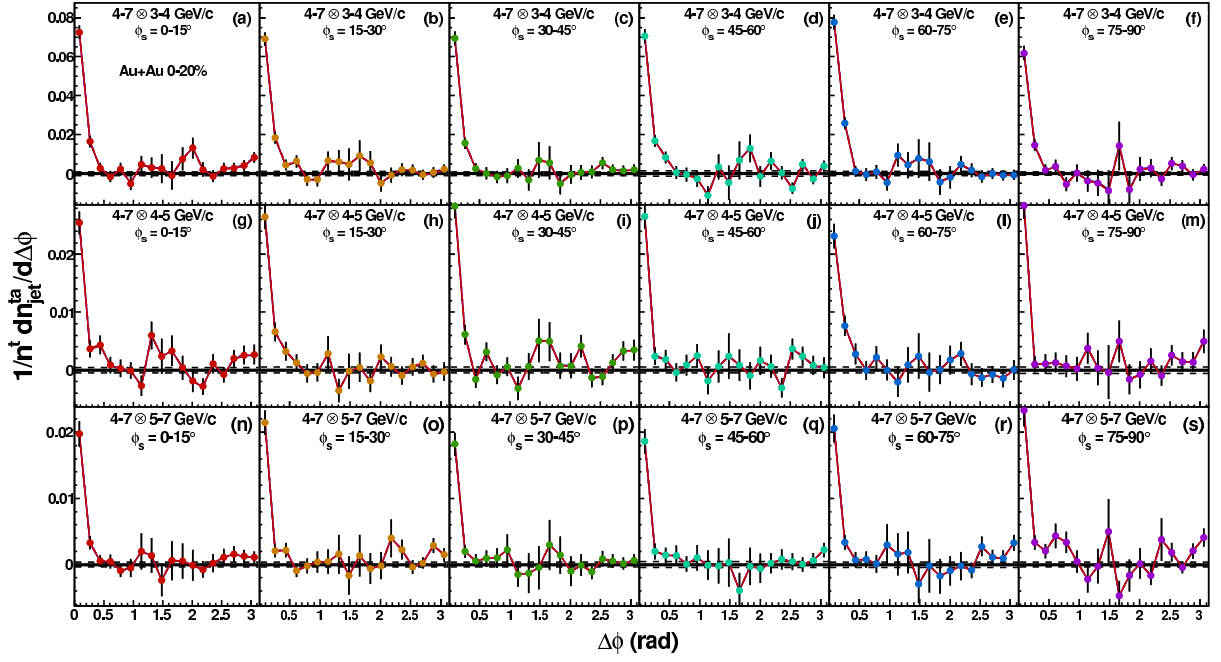}
  \caption{\label{fig:jfsall_00_20}
    (Color online) Per-trigger azimuthal jet yields in central 0--20\%
    collision for all trigger orientations with respect to the
    reaction plane in 15$^{\circ}$ selections from the most in-plane
    to the most out-of-plane proceeding from left to right.  Associated
    partner momentum selections from lower to higher $p^a_T$ are arranged
    from top to bottom.  Bars indicate statistical uncertainties.  Flow
    modulation systematic uncertainties are represented by bands
    through the points while flow normalization uncertainties are
    shown as dashed lines around zero.  
  } 
\end{figure*}

\begin{figure*}[ht]
  \includegraphics[width=0.90\linewidth]{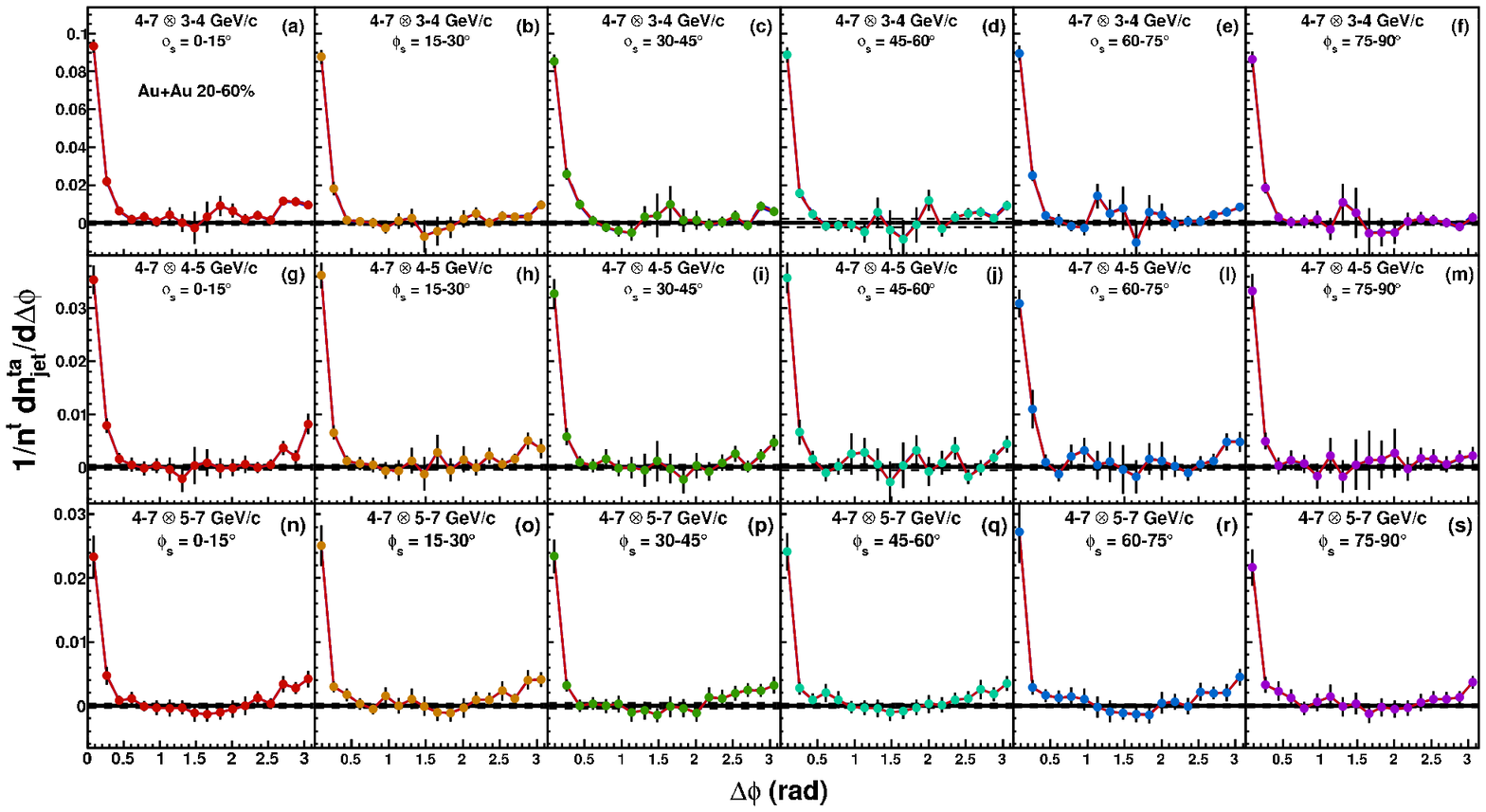}
  \caption{\label{fig:jfsall_20_60}
    (Color online) Per-trigger azimuthal jet yields in midcentral 20--60\%
    collision for all trigger orientations with respect to the
    reaction plane in 15$^{\circ}$ selections from the most in-plane
    to the most out-of-plane proceeding from left to right.  Associated
    partner momentum selections from lower to higher $p^a_T$ are arranged
    from top to bottom.  Bars indicate statistical uncertainties.  Flow
    modulation systematic uncertainties are represented by bands
    through the points while flow normalization uncertainties are
    shown as dashed lines around zero.  
  } 
\end{figure*}

\begin{figure*}[ht]
  \includegraphics[width=0.90\linewidth]{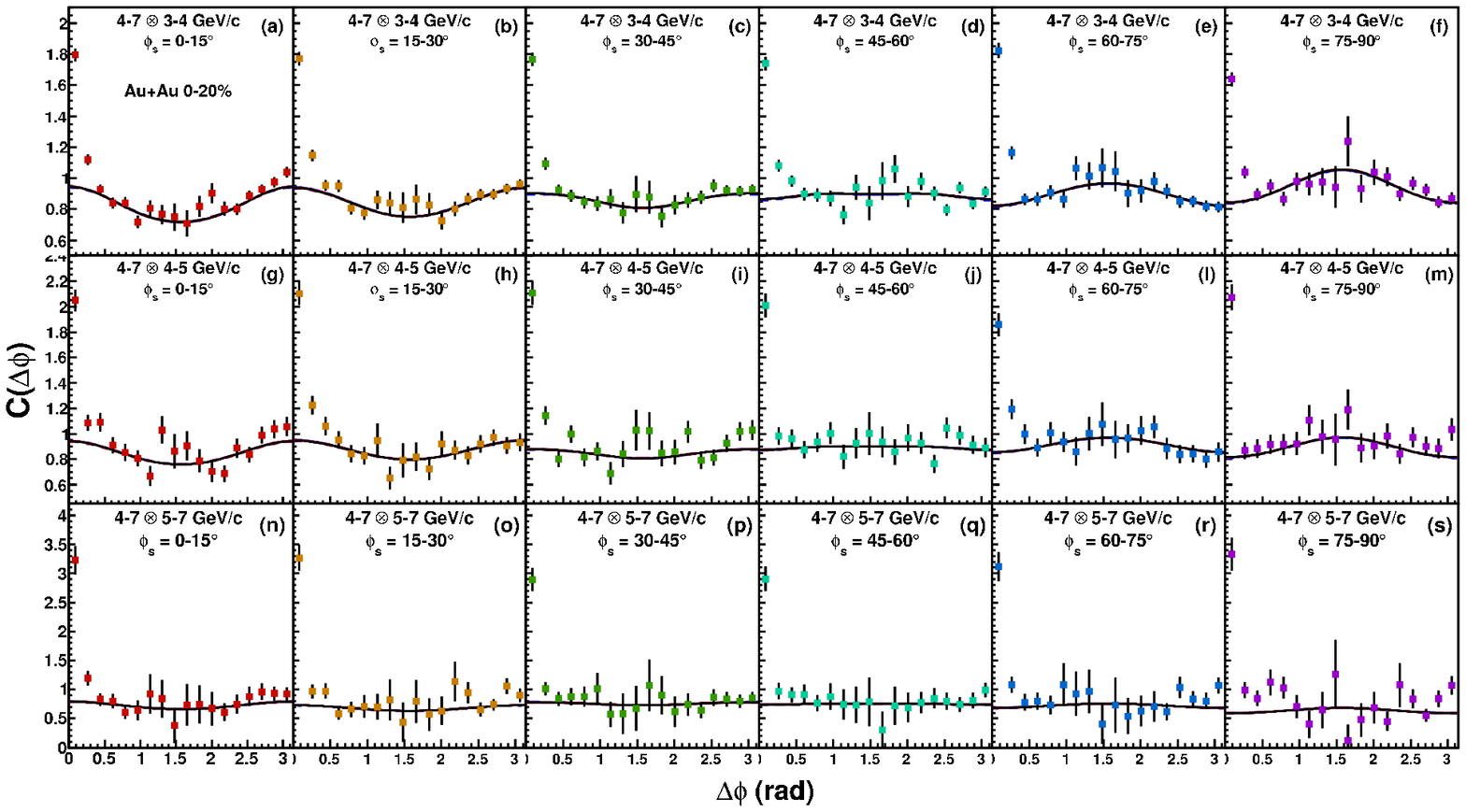}
  \caption{\label{fig:cfsall_00_20}
    (Color online) Correlation functions in 0--20\% central collisions
    for all trigger orientations with respect to the reaction plane in
    15$^\circ$ selections from most in-plane to most out-of-plane
    proceeding from left to right.  Associated partner momentum
    selections from lower to higher $p^a_T$ arranged from top to
    bottom.  Expected flow contributions are shown as solid curves (see
    text for details).   
  } 
\end{figure*}

\begin{figure*}[ht]
  \includegraphics[width=0.90\linewidth]{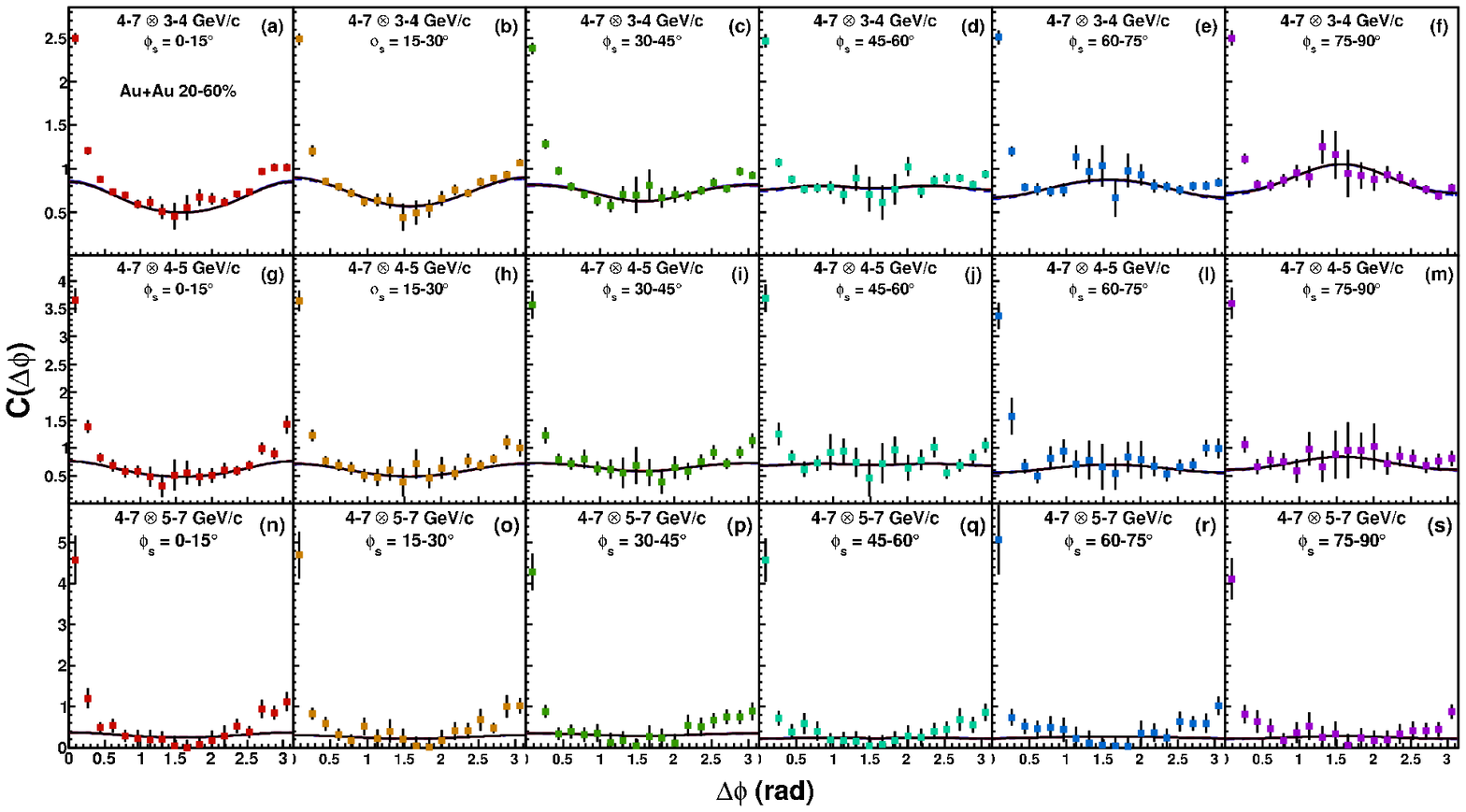}
  \caption{\label{fig:cfsall_20_60}
    (Color online) Correlation functions in 20--60\% central collisions
    for all trigger orientations with respect to the reaction plane in
    15$^\circ$ selections from most in-plane to most out-of-plane
    proceeding from left to right.  Associated partner momentum
    selections from lower to higher $p^a_T$ arranged from top to
    bottom.  Expected flow contributions are shown as solid curves (see
    text for details).   
  } 
\end{figure*}

The complete set of per-trigger yields and correlation functions used
as source material for the analysis of the dependence of the
away-side suppression on angle with respect to the reaction plane are
shown in Figs.~\ref{fig:jfsall_00_20}--\ref{fig:cfsall_20_60}.

\clearpage


\end{document}